\documentclass[3p]{elsarticle}

\usepackage{color}
\usepackage{amssymb, amsmath}
\usepackage{multirow}
\usepackage{url}

\usepackage{algorithm2e}

\usepackage{fancyhdr}
\usepackage{listings}

\journal{Parallel Computing, accepted 5/16/18}

\begin{document}

\thispagestyle{fancy}

\begin{frontmatter}

\title{Parallel Accelerated Custom Correlation Coefficient Calculations
for Genomics Applications\tnoteref{ornltitlenote}}

\tnotetext[ornltitlenote]{This manuscript has been authored by UT-Battelle,
LLC under Contract No. DE-AC05-00OR22725 with the U.S. Department of
Energy.  The United States Government retains and the publisher, by
accepting the article for publication, acknowledges that the United States
Government retains a non-exclusive, paid-up, irrevocable, world-wide
license to publish or reproduce the published form of this manuscript, or
allow others to do so, for United States Government purposes.  The
Department of Energy will provide public access to these results of
federally sponsored research in accordance with the DOE Public Access Plan
(http://energy.gov/downloads/doe-public-access-plan).
\vskip .05in
Accepted to {\it Parallel Computing}, 5/16/18. \copyright \ 2018. This manuscript version is made available under the CC-BY-NC-ND 4.0 license \url{http://creativecommons.org/licenses/by-nc-nd/4.0/}.
\vskip .05in
}


\author[ornladdress]{Wayne Joubert\corref{mycorrespondingauthor}}
\cortext[mycorrespondingauthor]{Corresponding author}
\ead{joubert@ornl.gov}
\author[ornladdress]{James Nance}
\author[umsl]{Sharlee Climer}
\author[ornladdress,utkaddress]{Deborah Weighill}
\author[ornladdress,utkaddress]{Daniel Jacobson}

\address[ornladdress]{
 Oak Ridge National Laboratory,
 1 Bethel Valley Road,
 Oak Ridge, TN 37831
}

\address[utkaddress]{
The Bredesen Center for Interdisciplinary Research and Graduate Education,
University of Tennessee, Knoxville,
444 Greve Hall, 821 Volunteer Blvd.
Knoxville, TN 37996-3394
}

\address[umsl]{
University of Missouri-St. Louis,
1 University Blvd. 311 Express Scripts Hall,
St. Louis, MO 63121-4400
}


\begin{abstract}

The massive quantities of genomic data being made available
through gene sequencing techniques are enabling breakthroughs
in genomic science in many areas such as medical advances
in the diagnosis and treatment of diseases.
Analyzing this data, however, is a computational challenge
insofar as the computational costs of the relevant algorithms
can grow with quadratic, cubic or higher complexity---leading
to the need for leadership scale computing.
In this paper we describe a new approach to calculations of
the Custom Correlation Coefficient (CCC)
between Single Nucleotide Polymorphisms (SNPs) across a population,
suitable for parallel systems equipped with graphics
processing units (GPUs) or Intel Xeon Phi processors.
We describe the mapping of the algorithms to accelerated processors,
techniques used for eliminating redundant calculations due to symmetries,
and strategies for efficient mapping of the calculations
to many-node parallel systems.
Results are presented demonstrating high per-node performance and
near-ideal parallel scalability
with rates of more than nine quadrillion ($9\times 10^{15}$)
elementwise comparisons achieved per second with the
latest optimized code on the ORNL Titan system,
this being orders of magnitude faster than rates achieved using
other codes and platforms as reported in the literature.
Also it is estimated that as many as 90 quadrillion
($90\times 10^{15}$) comparisons per second may be achievable
on the upcoming ORNL Summit system, an additional 10X performance increase.
In a companion paper we describe corresponding techniques applied
to calculations of the Proportional Similarity metric
for comparative genomics applications.

\end{abstract}



\end{frontmatter}



\section{Introduction}
\label{section:introduction}
Computation of the mathematical relationships between pairs
of vectors is required in many science domains.
In the field of genomics,
the Custom Correlation Coefficient (CCC)~\cite{ccc}
was developed to calculate the correlation between mutations,
or Single Nucleotide Polymorphisms (SNPs), across a population of individuals.
This can be used to identify groups of SNP alleles which tend to co-occur
in a population and consequently can be used to find combinations
of SNP alleles which associate with certain phenotypes,
such as a disease phenotype~\cite{climer2014allele}.
CCC also takes into account genetic heterogeneity and finds
correlations between SNP alleles which co-occur in portions of the population,
not requiring co-occurrence across the whole population.

The effectiveness of CCC has been demonstrated in diverse domains. It has been used to identify genetic patterns exhibiting highly significant associations with both hypertensive heart disease~\cite{ccc} and psoriasis~\cite{climer2014allele}. In another study, CCC was utilized to reveal genetic coadaptation between vitamin D receptor and skin color genes, suggesting parallel selective responses to environmental transitions as humans ventured out of Africa~\cite{tiosano2016latitudinal}. Interestingly, the CCC analysis of HapMap~\cite{frazer2007second} data revealed a large-scale genetic pattern on human chromosome 14~\cite{climer2015human}. This discovery suggests that two completely divergent evolutionary paths rapidly progressed in our past, presumably achieving the shared goal of enhancing gephyrin, a gene that is vital for signal transmissions in the human brain. Note that the HapMap data are arguably the most extensively studied SNP data available, yet this pattern was not previously identified by others---demonstrating CCC's ability to identify combinatorial patterns of correlated SNP alleles within genome-wide data that are missed by other methods. 

The explosive growth in genomic data has opened
unprecedented levels of opportunity for addressing fundamental
questions of importance in genomic sciences.
However, the requisite computational methods are expensive,
insofar as the computational complexity of general pairwise
vector comparison methods is quadratic in the number of vectors,
and the complexity is even higher for methods
comparing three or more vectors at a time
as is required in some cases.
To perform large scale studies, it has thus become necessary to
use high performance leadership computing systems
possessing thousands of compute nodes equipped
with advanced hardware such as accelerated manycore
processors.

In this paper we describe new approaches to performing CCC
calculations on leadership class systems equipped with GPUs.
We describe implementations of CCC methods which
attain high absolute performance on GPUs,
use asynchronous methods to overlap operations,
avoid the performance penalty of
performing redundant and unnecessary computations,
and exhibit near-ideal scaling to thousands of compute
nodes.


Substantial recent work has focused on the problem of
comparing pairs, triples or larger subsets of a set of vectors
efficiently by means of advanced computational methods
such as the use of parallelism, accelerated GPU or Intel Xeon Phi
processing, or both.
A broad overview of epistasis detection in comparative genomics
including computational issues pertaining to parallelism and GPU acceleration
is given in \cite{wei}.
The GBOOST code, discussed in
\cite{gboost},
is a gene-gene interaction code for 2-way
studies optimized for single GPUs using encoding of gene data into bit
strings with avoidance of redundant computations;
Wang et al.\ \cite{gwisfi}
describes GWISFI, a single-GPU code for 2-way GWAS calculations.
Gonzalez-Dominguez et al. \cite{gonzalez}
develops a UPC++ code for gene-gene interaction studies
for small numbers of GPUs and Intel Phi processors
exploiting vector hardware and hardware population count instructions.
Gonzalez-Dominguez and Schmidt \cite{gonzalez2}
considers 3-way interactions on a node with 4 GPUs.
Solomonik et al.\ \cite{solomonik}
develops parallel tensor computation methods,
structurally similar to 3-way metrics computations, with
particular attention to avoiding redundant computations;
however, the work does not consider GPUs or shaping of the
computational regions to accommodate processors with long vector lengths.
Haque et al.\ \cite{pande}
discusses similarity metric calculations for chemical informatics
applications
on single GPUs using space filling curve methods and hardware
population count instructions;
it recognizes the correspondence of these calculations to
BLAS-3 matrix-matrix product computations and pays close
attention to optimizing memory accesses.
Wang et al.\ \cite{cloud} considers 2-way studies on compute clouds using MapReduce
on conventional CPUs.
Yang et al.\ \cite{pawsey}
adapts existing packages to perform 2-way CPU and GPU
studies and 3-way CPU studies on as many as 200 cores in
parallel.
Goudey et al.\ \cite{goudey}
performs k-way GWAS studies for arbitrary k with consideration of
load balancing and elimination of redundancies
on a 4096-node IBM Blue Gene/Q system;
results for a single GPU are also presented.
Luecke et al.\ \cite{weeks}
performs 2-way analyses on up to 126 nodes of the Intel
Phi-based Stampede system (cf. \cite{weeks2}).
Koesterke et al.\ \cite{stanzione}
considers 2-way computations on thousands of compute cores
with good scalability and good absolute performance on
conventional CPUs.
Finally, recent work in \cite{fb}
considers $k$-selection similarity search methods with applications
to image data with results for small numbers of GPUs;
that work however focuses primarily on the $k$-selection problem for
nonexhaustive inexact similarity search,
a different problem from what is considered here.


This work is to our knowledge the first
successful effort
to combine all the needed elements for performing large-scale 2-way
and 3-way vector comparison studies on leadership-class
systems, including:
high performance usage of accelerated processors,
effective use of deep memory hierarchies,
avoidance of unneeded redundant computations,
effective scaling to thousands of compute nodes,
and algorithm structuring to enable efficient I/O.

The remainder of this paper is organized as follows.
After describing the 2-way and 3-way CCC methods
in Section~\ref{section:ccc},
we describe the techniques used to map these methods to GPUs
and other manycore accelerated processors in
Section~\ref{section:gpus}.
Then we describe the parallelization techniques applied to
these methods
in Section~\ref{section:parallel},
followed by implementation details in
Section~\ref{section:implementation}.
Computational results on the 27 petaflop
Oak Ridge National Laboratory (ORNL) Cray XK7 Titan system
are presented in Section~\ref{section:results}.
Discussion of future work is presented in
Section~\ref{section:discussion},
and conclusions are given in
Section~\ref{section:conclusions}.

For additional discussion of shared topics
pertaining to the algorithms and their implementations,
it is advised that this paper be read in tandem with
the companion paper~\cite{companionczek}.


\section{The Custom Correlation Coefficient}
\label{section:ccc}

\subsection{The 2-way metric}

We assume a set of $n_v$ vectors of length $n_f$ elements
$\{v_i\}_{i=1}^{n_v}$ with $v_i\in (S_2)^{n_f}$ and
$v_i=\{v_{i,q}\}_{q=1}^{n_f}$.
Here $S_2 = S\times S = S^2$ where $S=\{0, 1\}$, thus each vector entry
$v_{i,q}$ is itself a vector with two entries
$\{(v_{i,q})_r\}_{r=1}^2$ taken from $S$.
In practice, the $v_i$ are SNPs each of whose entries
$v_{i,q}\in S_2$ represents a pair of alleles, with a possibly
different allele interpretation for each column $i$, and $n_f$
is the number of samples or population size.

For $a, b\in S$ define the indicator function $\chi_a$ by
$\chi_a(b)=1$ if $a=b$, otherwise $0$.
Let $\rho_{i,q}(a) = \sum_r \chi_a((v_{i,q})_r)$,
the count of entries with the value $a$ in $v_{i,q}$.
The \emph{frequency} of allele $a$ for SNP $v_i$ is then
\begin{align}
f_i(a) = \frac{1}{2n_f}\sum_{q=1}^{n_f}\rho_{i,q}(a).
\end{align}
Clearly $f_i(0)+f_i(1)=1$.
Also let $\rho_{i,j,q}(a,b)=\rho_{i,q}(a) \cdot \rho_{j,q}(b)$ and
\begin{align}
f_{i,j}(a,b) = \frac{1}{4n_f}\sum_{q=1}^{n_f}\rho_{i,j,q}(a,b).
\end{align}
Note $\sum_{a, b \in S} f_{i,j}(a,b)=1$.
Then the 2-way CCC comparison for $a, b \in S$ assuming a fixed constant
$\gamma=2/3$ is
\begin{align}
CCC_{i,j}(a,b) = f_{i,j}(a,b)
(1-\gamma f_i(a)) (1-\gamma f_j(b)) .
\end{align}

The functions $f_{i,j}()$, and thus $CCC_{i,j}()$, are symmetric
in $i$ and $j$.
Thus to compute all unique values
$\{f_{i,j}(a,b)\}_{i,j,a,b}$ for distinct $i$ and $j$
requires $4 n_f n_v (n_v-1)/2 = O(n_f n_v^2)$ operations.
On the other hand, $\{f_i(a)\}_{i,a}$
requires only $2 n_f n_v = O(n_f n_v)$ operations.
Due to its greater computational cost, the efficient
calculation of $\{f_{i,j}(a,b)\}$ will be the chief focus of
this work.

\begin{figure}[ht]
\centering
\includegraphics*[height=2in]{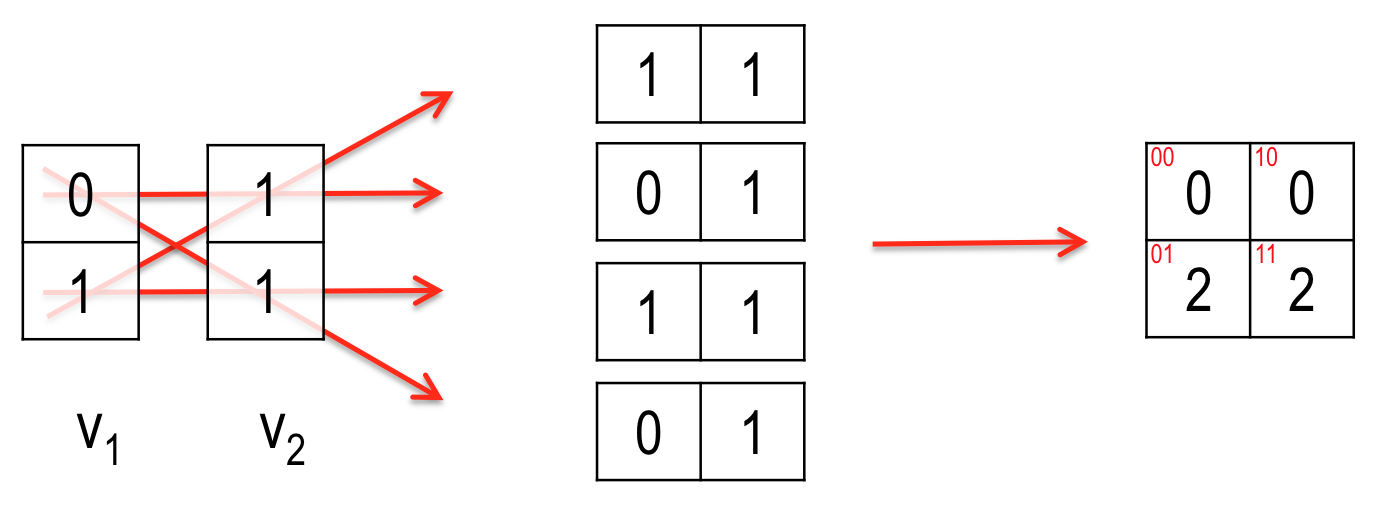}
\caption{2-way CCC calculation example. Left: two vectors of
length 1, with each entry a 2-vector. Center: enumeration of all
pairings of entries. Right: tallying of counts of each
pairing type.}
\label{fig:ccc-2way-example}
\end{figure}

An interpretation of the $f_{i,j}$ component of the
2-way CCC calculation is shown in
Figure~\ref{fig:ccc-2way-example}.
Here we let $n_f=1$ and $n_v=2$.
For the first entry of $v_1$ and of $v_2$,
each containing two binary entries as shown, all four pairings of
the left two entries and the right two entries
are selected and enumerated, with four resulting tuple values, each
taken from a set of four possible combinations $(0,0)$,
$(0,1)$, $(1,0)$ and $(1,1)$.
These tuples are then tallied by value to count the
frequency of each possible tuple.
For the general case of $n_f>1$, the additional entries
are handled in the same
way, with the counts of each pairing tally summed into the
result table shown at the right.

\subsection{The 3-way metric}

The previous section described CCC for evaluating pairs of SNPs;
next we expand this to the evaluation of SNP trios.
The 3-way CCC comparison for $a, b, c\in S$ and vectors
$v_i, v_j, v_k$ and a fixed constant $\gamma$ is defined by
\begin{align}
CCC_{i,j,k}(a,b,c) = f_{i,j,k}(a,b,c) (1-\gamma f_i(a)) (1-\gamma f_j(b)) (1-\gamma f_k(c)) .
\end{align}
Here, 
\begin{align}
f_{i,j,k}(a,b,c) =
\frac{1}{8n_f}\sum_{q=1}^{n_f}\rho_{i,j,k,q}(a,b,c)
\end{align}
for
$\rho_{i,j,k,q}(a,b,c)=\rho_{i,q}(a) \cdot \rho_{j,q}(b) \cdot \rho_{k,q}(c)$.

Due to symmetries in $i$, $j$ and $k$,
only $8 n_v (n_v-1) (n_v-2)/6$
unique values of
$\{f_{i,j,k}(a,b,c)\}_{i,j,k,a,b,c}$ need be computed
for distinct $i$, $j$ and $k$,
requiring $8 n_f n_v (n_v-1)(n_v-2)/6 = O(n_f n_v^3)$ operations.
The dominance of this computational expense over the
calculation of the $f_i(a)$ values makes this the primary
focus of attention.

\begin{figure}[ht]
\centering
\includegraphics*[height=2in]{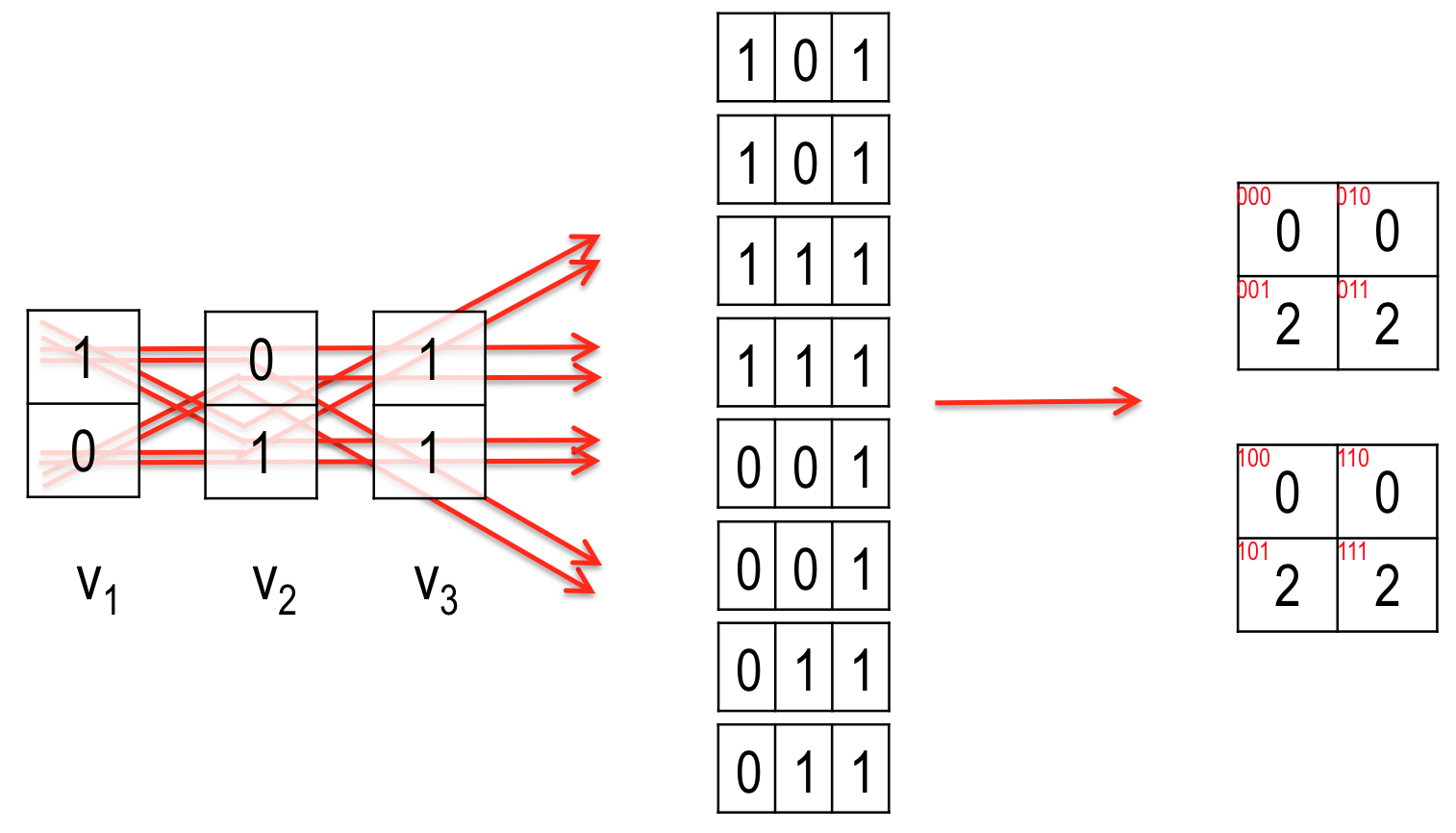}
\caption{3-way CCC calculation example. Left: three vectors of
length 1, with each entry a 2-vector. Center: enumeration of all
combinations of entries. Right: tallying of counts of each
combination type.}
\label{fig:ccc-3way-example}
\end{figure}

Figure~\ref{fig:ccc-3way-example} gives an interpretation,
for $n_f=1$ and $n_v=3$.
Again, each vector has one entry
which is itself a 2-vector of binary entries.
All eight combinations of vector entries are sampled, and the
counts of these triples are tallied into a table whose
entries correspond to the eight possible combinations of
three binary values.

\section{Mapping to manycore processors}
\label{section:gpus}

\subsection{The 2-way metric}

Let us define $B=V^T \circ_2 V$
to satisfy $(B_{i,j})_{a,b} = f_{i,j}(a,b)$,
so that each entry of $B$ is a $2\times 2$ tally table.
It can be observed that the basic structure of the computation of the
$f_{i,j}(a,b)$ values---namely, the summing of values derived
from corresponding entries of each pair of vectors---is
identical to the structure of the general matrix-matrix
product computation (GEMM), with the pairwise multiplication of
scalar vector elements in the GEMM replaced with a tally into a
2$\times$2 table.
To solve this efficiently on manycore compute nodes, we
follow the approach of~\cite{companionczek},
adapting the GEMM kernel from
an optimized dense linear algebra library,
in this case MAGMA\cite{magma} targeting GPUs,
to accelerate this operation.

In~\cite{companionczek} we implemented the Proportional Similarity
metric by replacing the scalar multiplication operation
of the GEMM with the
operation of taking the minimum of two scalar values.
Here however we have a tally operation which requires special
considerations to implement efficiently.
For CCC, since every entry of $v_i$ is composed of only two bits,
significant savings in computation time as well as storage
is possible by packing the bits of many vector entries into
a single compute word.
We use the MAGMA double complex ZGEMM operation,
for which each MAGMA vector entry 
is composed of two 64-bit words, so that a
single ZGEMM vector entry can hold 64 $v_i$ entries.
For fixed $i$ and $j$, four $f_{i,j}(a,b)$ (integer)
values must then be accumulated.
To fit this into a result value of two 64-bit doubles
for the ZGEMM,
we assign 25 bits of the mantissa of each of the two floating point
numbers to store each result.
This allows $n_f$ to be as large as $2^{23}-1 = 8,388,607$,
without loss of precision from overflow of the mantissa
or mixing of the
two integer components stored in each word.
This limit is far above the typical requirements
for the targeted calculations.
The resulting modified ZGEMM operation we refer to here as mGEMM2.

For each pair of corresponding double complex vector entries
for which a tally is performed, it is efficient to
use bitwise operations.
In the implementation,
to accumulate to tally targets $(0,0)$, $(0,1)$, $(1,0)$ and $(1,1)$,
corresponding words are operated on by
left and right bit shifts and bitwise OR and AND operations
to obtain words for
which each bit is set to 1 if the original pair of bits
corresponds to the respective tally table entry and 0 otherwise.
Then the CUDA intrinsic {\tt{}\_\_popcll()} corresponding to
a hardware population count operation is used to sum
the number of bits of the word set to 1 and accumulate the
result.
The population count operation is
supported in hardware for some
conventional x86 CPUs as well; if not, it can
typically be implemented with small numbers of machine
instructions; see for example~\cite{hakmem},
\cite{hackersdelight}.
In the implementation, care must be taken for the case when
$n_f$ is not a multiple of 64 so that the tallies
corresponding to the zero padding are corrected for.

Though we do not pursue the topic here, it is likely that
libraries optimized to other processor architectures, such as
PLASMA~\cite{plasma}, BLIS~\cite{blis} and OpenBLAS \cite{openblas}
would provide similar opportunities for
high performance for conventional processors and Intel Xeon Phi.


\subsection{The 3-way metric}

\begin{table}
\begin{center}
\caption{Values of $X_{j,\xi}$}
\label{tab:xjc}
\vspace{.05in}
\begin{tabular}{cc|ccc}
\hline
$V$  &   $v_j$   &  $X_{j,1}$  &  $X_{j,2}$  &  $X_{j,3}$  \\
\hline
0,0  &  0,0  &   0,0     &   1,0     &   1,0     \\
0,0  &  0,1  &   0,1     &   1,0     &   1,0     \\
0,0  &  1,1  &   1,1     &   1,0     &   1,0     \\
\hline
0,1  &  0,0  &   1,0     &   0,0     &   1,0     \\
0,1  &  0,1  &   1,0     &   0,1     &   1,0     \\
0,1  &  1,1  &   1,0     &   1,1     &   1,0     \\
\hline
1,1  &  0,0  &   1,0     &   1,0     &   0,0     \\
1,1  &  0,1  &   1,0     &   1,0     &   0,1     \\
1,1  &  1,1  &   1,0     &   1,0     &   1,1     \\
\end{tabular}
\end{center}
\end{table}

The 3-way method cannot be mapped directly to a modified GEMM
framework using the same approach,
insofar as in this case the tally table entries would
need to be reduced to 12
bits to fit into two double precision words of the ZGEMM,
limiting $n_f$ to size 4095 or less, inadequate for many problems.
Thus an alternative approach is needed.

To solve this problem, the basic approach taken here is a
two-step process.
First, for given $V$ and a fixed column $v_j$ we construct modified
matrices $X_{j,\xi}$, $1\leq \xi\leq 3$,
each of the same dimension as $V$.
Second, we calculate $B_\xi = X_{j,\xi}^T \circ_3 V$ for an
operator $\circ_3$ representing a modified 2-way calculation
from which it is possible to reconstruct the 3-way tally values
for $v_i$, $v_j$ and $v_k$, for $1\leq i, k\leq n_v$.

For the first step,
the entry $(X_{j,\xi})_{q,p}$ is formed from the entries
$v_{j,q}$ and $(V)_{q,p}$ according to the rules shown in
Table~\ref{tab:xjc}.
The table gives values for the $n_f=1$ case; for the
general case, corresponding rows of $v_j$ and $V$
are handled likewise.
Here, for convenience the entries of $v_j$ and $V$
that have the value $(1,0)$ are omitted since they
have an identical result on the calculation as
$(0,1)$ insofar as the CCC result tally values do not
differentiate between these cases.

The aim here is,
for each column of $V$ (with $(1,0)$ entries mapped to $(0,1)$),
to form the corresponding columns of $X_{j,1}$, $X_{j,2}$ and $X_{j,3}$
by taking the corresponding entry of $v_j$
(again replacing $(1,0)$ with $(0,1)$)
if the corresponding entry of the $V$ column
equals $(0,0)$, $(0,1)$ and $(1,1)$ respectively, and
the ``null'' indicator value of $(1,0)$ otherwise.
We refer to this modified GEMM operation as mGEMM3.


After $\{X_{j,\xi}\}_\xi$ are formed, the matrices
$B_{j,\xi} = X_{j,\xi}^T \circ_3 V$
are calculated similarly to the previously described 2-way method
based on modified GEMMs.
Here the operator $\circ_3$ represents the same tally
process as used for the 2-way method described above except that
$X_{j,\xi}$ values equal to the null indicator $(1,0)$ are discarded
and not used in the tally.

It is then straightforward to construct the 3-way
tallies from the constituent 2-way tallies $\{B_{j,\xi}\}_\xi$.
The matrix $B_{j,1}$ contributes tallies for entries of
$V$
equal to $(0,0)$, and similarly
$B_{j,2}$ for $(0,1)$ and $B_{j,3}$ for $(1,1)$.
For each $B_{j,\xi}$, $1\leq\xi\leq 3$, the four 2-way tally results
for each vector triple must be mapped to the appropriate
eight table entries for the 3-way tally associated with this
vector triple.
The mapping of contributions for each $\xi$ is given by
$$
f_{i,j,k}(0,0,0) = 2((B_{j,1})_{(i,k)})_{(0,0)} + ((B_{j,2})_{(i,k)})_{(0,0)}
,$$
$$
f_{i,j,k}(0,0,1) = 2((B_{j,1})_{(i,k)})_{(0,1)} + ((B_{j,2})_{(i,k)})_{(0,1)}
,$$
$$
f_{i,j,k}(0,1,0) = 2((B_{j,1})_{(i,k)})_{(1,0)} + ((B_{j,2})_{(i,k)})_{(1,0)}
,$$
$$
f_{i,j,k}(0,1,1) = 2((B_{j,1})_{(i,k)})_{(1,1)} + ((B_{j,2})_{(i,k)})_{(1,1)}
,$$
$$
f_{i,j,k}(1,0,0) = 2((B_{j,3})_{(i,k)})_{(0,0)} + ((B_{j,2})_{(i,k)})_{(0,0)}
,$$
$$
f_{i,j,k}(1,0,1) = 2((B_{j,3})_{(i,k)})_{(0,1)} + ((B_{j,2})_{(i,k)})_{(0,1)}
,$$
$$
f_{i,j,k}(1,1,0) = 2((B_{j,3})_{(i,k)})_{(1,0)} + ((B_{j,2})_{(i,k)})_{(1,0)}
,$$
$$
f_{i,j,k}(1,1,1) = 2((B_{j,3})_{(i,k)})_{(1,1)} + ((B_{j,2})_{(i,k)})_{(1,1)}
.$$


The 3-way calculation thus requires three modified GEMM operations,
compared to one for the 2-way method.
Insofar as each 3-way vector triple tally requires eight values
compared to four for each 2-way pair,
and the path length (Figure~\ref{fig:ccc-3way-example})
is three for the 3-way method compared to two
for the 2-way method (Figure~\ref{fig:ccc-2way-example}),
we believe it is near-optimal
for the 3-way method computed using bitwise arithmetic
to require roughly 3X
the work of the 2-way method, as is the case in this implementation.


\section{Multi-node parallelism}
\label{section:parallel}

The details of the parallel implementation are essentially identical
to those for the Proportional Similarity metric as described in detail
in~\cite{companionczek}.
Here we give the main ideas in summary form.

Parallelism is obtained by decomposing
both the $n_f$ problem dimension
and the $n_v$ problem dimension across $n_{pf}$ and $n_{pv}$
nodes respectively, resulting in $n_{fp}\times n_{vp}$
elements per node.
Here for convenience we assume one GPU per node, though
the implementation does not require this.
A third axis, $n_{pr}$, is used to apply additional parallelism
across the computation of the result values.

The $n_{pf}$ parallelism axis requires a parallel reduction operation.
The $n_{pv}$ axis requires carefully scheduled point-to-point
communications so that every vector can be compared against
every other vector.

For the 2-way method, the result values form a 2-D square
matrix of values with an imposed decomposition into smaller
square blocks due to the $n_{pv}$ parallelism.
To avoid computing redundant values
resulting from symmetry of the matrix,
results are computed only for a block circulant subset of
the blocks.
Each $n_{pv}$-parallelism compute node is responsible for computing
a block row of this matrix.
The $n_{pr}$ parallel axis is used to parallelize the computation
of the blocks of this block row.

For the 3-way method, the result values make up a
cube-shaped 3-D array of values, implicitly
decomposed into smaller cubes by the $n_{pv}$ parallel
decomposition.
A scheme is implemented so that only a subset or ``slice'' of values
is calculated for each block, this subset chosen to
represent each of the unique values in the result exactly
once.
Each $n_{pv}$-parallelism compute node is responsible for computing
a 2-D slab of the results cube.
The $n_{pr}$ axis of parallelism is deployed so that
the blocks in this slab can be computed in parallel.

The 3-way method allows an additional setting, $n_{st}$,
which allows the metrics computation to be broken into
multiple stages.
To reduce the main memory storage costs for the metrics values,
a run can be performed for which only a single stage of results is
computed.

Asynchronous methods are used to overlap GPU computations,
CPU computations, communications and transfers of data to and from the GPUs.


\section{Implementation}
\label{section:implementation}

The algorithms described here are implemented in
the CoMet parallel genomics code.
This code is written in C++, compiles with the GNU compiler suite
and depends on MPI, CUDA and the modified versions of the MAGMA library.
GNU Make and CMake are used for build management, and googletest
is used for unit testing.
The clang-format source code tool from the clang compiler package is used for
source code formatting, and Git is used for repository management.

OpenMP CPU threading is used to accelerate the parts of the
computation that are not ported to the GPU
by mapping execution to multiple CPU
cores on the node;
when possible, the CPU work is also hidden under the
asynchronously launched GPU kernels to improve performance.

For making comparisons, each method has a reference (CPU-only) version,
a (possibly optimized) CPU version, and a GPU version.
A set of synthetic reference test cases is implemented for
testing, designed to give the exact same bit-for-bit result
for all code versions and for all parallel decompositions.
Two types of synthetic problem are implemented:
a version for which each vector entry is set to a randomized value,
and a second version with randomized placement of entries specifically
chosen so that the correctness of every result value can be
verified analytically.
A checksum feature using extended precision integer
arithmetic computes a bit-for-bit exact checksum of
computed results to check for errors when using
synthetic inputs.

The code can be compiled under single or double precision.
The precision setting for the CCC case affects only the accuracy
of the calculation of the $f_i()$ values for the denominators;
for the numerator computations, as described earlier
the relevant computations are performed
with double complex data types operated on primarily with bitwise operations.

To modify MAGMA as needed for the algorithms, it is
necessary to modify the two files in the MAGMA package
{\tt{}magmablas/gemm\_stencil.cuh} and
{\tt{}magmablas/gemm\_stencil\_defs.h}.
In particular, the macro definition for ``{\tt{}fma}'' defining the
fused multiply accumulate must be changed to make use of the
appropriate tally operation.


\section{Computational results}
\label{section:results}

\subsection{Overview}

Experiments are performed on the ORNL Titan Cray XK7
system.
Titan is composed of 18,688 compute nodes each equipped with
an AMD Interlagos 16 core CPU and an NVIDIA Kepler K20X GPU
connected via a PCIe-2 bus.
The K20X GPU has peak single/double precision flop rate of
3,935/1,311 GF and peak memory bandwidth of 250 GB/sec.
Each node contains 32 GB main memory and 6 GB GPU memory.

The software versions used are
Cray OS version 5.2.82,
Cray Programming Environment 2.5.13,
GCC  4.9.3,
MAGMA 1.6.2
and
CUDA toolkit 7.5.18-1.0502.10743.2.1.
For large node counts, it is in some cases necessary to set the
environment variable {\tt{}APRUN\_BALANCED\_INJECTION}
to values such as 63 or 33 to avoid throttling of the
communication network resulting from the
algorithms' communication patterns
and causing performance loss.

The primary use of the code is to solve very large
problems not previously solvable; thus weak scaling
behavior,
for which the work per node is kept roughly constant as
compute node count is increased, is the primary focus.

GPU-enabled runs are executed
with one MPI rank and one GPU per Titan node.
Reported execution times do not include I/O.
The source code execution path for the algorithm
is identical independent of
the actual values contained in the input vectors;
thus we expect performance for the synthetic datasets used
here to be essentially identical to performance with actual
genomics data.


\subsection{Single GPU kernel performance}

We first evaluate the raw performance of the modified GEMM kernels
in comparison with the standard GEMM.
We test the mGEMM2 and mGEMM3 operations described above,
used for the 2-way and 3-way methods respectively.
We use
$n_v=10,240$ vectors of length $n_f=393,216$ 2-bit values
corresponding to 6,144 double complex values.
Timings are compared against the standard ZGEMM for matrices
of the same size in memory.
Kernel times are taken from the CUDA Profiler
and include kernel time only, without transfer or CPU times.

Results are shown in Table~\ref{tab:1gpu}.  Raw timings are
shown as well as normalized times representing the time per
pair of elements operated on,
where an element is interpreted
to be a double complex value except for the modified GEMM
cases in which case it is a 2-bit value.
The cost of mGEMM3 is higher than that of mGEMM2 due to additional
integer masks and other operations needed.
The modified GEMMs have 64X higher density of vector elements per
double complex value, thus suggesting a higher throughput rate than ZGEMM
is possible; however, instead of four fused multiply-add
FMA operations per pair in the ZGEMM case,
a significant number of bitwise operations such as shifts,
masks, bitwise operations and population counts
are required per element pair.
Thus the modified GEMM normalized rates exceed the ZGEMM theoretical
peak by a smaller value than expected.
A side effect is that computational intensity is extremely
high, suggesting potentially high processor utilization
and favorable opportunities to hide communications and GPU data transfers
under computations.
The MAGMA ZGEMM rates are somewhat less than those of cuBLAS
since the former is more targeted toward smaller cases
required by other MAGMA operations.
Possible further optimizations of the bitwise operations
of the modified GEMMs
will be a topic of further study.

\begin{table}
\begin{center}
\caption{Kernel performance for single GPU case}
\label{tab:1gpu}
\vspace{.05in}
\begin{tabular}{cccc}
                        & time    & time per     & element pairs \\
                        & (sec)   & element pair & per second \\
\hline 
mGEMM2                  & 163.901 &  1.716E-12   & 5.828e11 \\
mGEMM3                  & 217.294 &  2.453E-12   & 4.077e11 \\
ZGEMM, MAGMA            &   6.998 & 10.863e-12   & 0.921e11 \\
ZGEMM, cuBLAS           &   4.493 &  6.974e-12   & 1.434e11 \\
ZGEMM, theoretical peak &   3.931 &  6.102e-12   & 1.639e11 \\
\hline 
\end{tabular}
\end{center}
\end{table}



\subsection{Performance model}

It is desirable to model algorithm performance in order to
evaluate expected performance and also to give guidance
regarding selection of tuning parameters.
We assume here that mGEMM2 and mGEMM3 sizes are large enough to hide
communications, GPU data transfers and CPU computations.

For the 2-way case, we define $\ell$, the ``load,''
to denote the number of blocks assigned to each node.
Then the execution time of the algorithm is estimated by
$$ t = t_C + t_{T,V} + \ell \cdot t_{G,2} + t_{T,M} + t_{CPU} , $$
where $t_C$ is the time for communicating $n_{fp} n_{vp}$ vector elements
per node for a parallel step,
$t_{T,V}$ the time to transfer $n_{fp} n_{vp}$ vector elements
to the GPU for a step,
$t_{T,M}$ the time to transfer $n_{vp}^2$ metrics values
from the GPU per step,
$t_{CPU}$ the time for denominator and quotient calculations
per step
and $t_{G,2}$ the time for an mGEMM2 computation.
The non-mGEMM2 times are included here to account
for asynchronous pipeline startup and drain.
It is evident that maximizing $\ell$ (by limiting $n_{pr}$)
makes it possible to approach peak mGEMM2 performance.
mGEMM2 rates are determined empirically; the
goal is to make the matrix dimensions $n_{fp}$ and
$n_{vp}$ for the mGEMM2 computation
 as large as possible to maximize mGEMM2 efficiency.
This suggests for a given problem it is desirable to reduce $n_{pv}$ and
$n_{pf}$ until CPU or GPU memory is filled.

For the 3-way case, we again define the load $\ell$, here
representing the number of block slices computed by a node.
Each slice is computed by a GPU pipeline of
$3(n_{vp}/6)/n_{st}$ mGEMM3 steps corresponding to the required
three mGEMM3 operations to form each 3-way result.
The execution time of the algorithm is estimated by
$$ t = t_C + t_{T,V} + \ell \cdot
 [ 3((n_{vp}/6)/n_{st})t_{G,3} + t_{T,V} + t_{T,M} + t_{CPU} ] . $$
Here mGEMM3 performance is approached by increasing $\ell$ and
$n_{vp}$, and decreasing $n_{st}$, subject to memory
constraints.
Similarly to the 2-way case, $n_{fp}$ and $n_{vp}$ should be
maximized in order to maximize mGEMM3 performance.


\subsection{2-way results}

For 2-way weak scaling results we set
$n_f=358,000$
elements per vector and
$n_{vp}=n_v / n_{pv}=4,096$ vectors per node.
We set the load $\ell=25$ and set
$n_{pr} = \lceil \lceil n_{pv} / 2 + 1 \rceil / \ell \rceil$ and $n_{pf}=1$.
The 2-way test runs are executed on up to
17,955 of Titan's 18,688 compute nodes, or 96.1\% of the system.

\begin{figure}[ht]
\centering
\includegraphics*[height=1.5in]{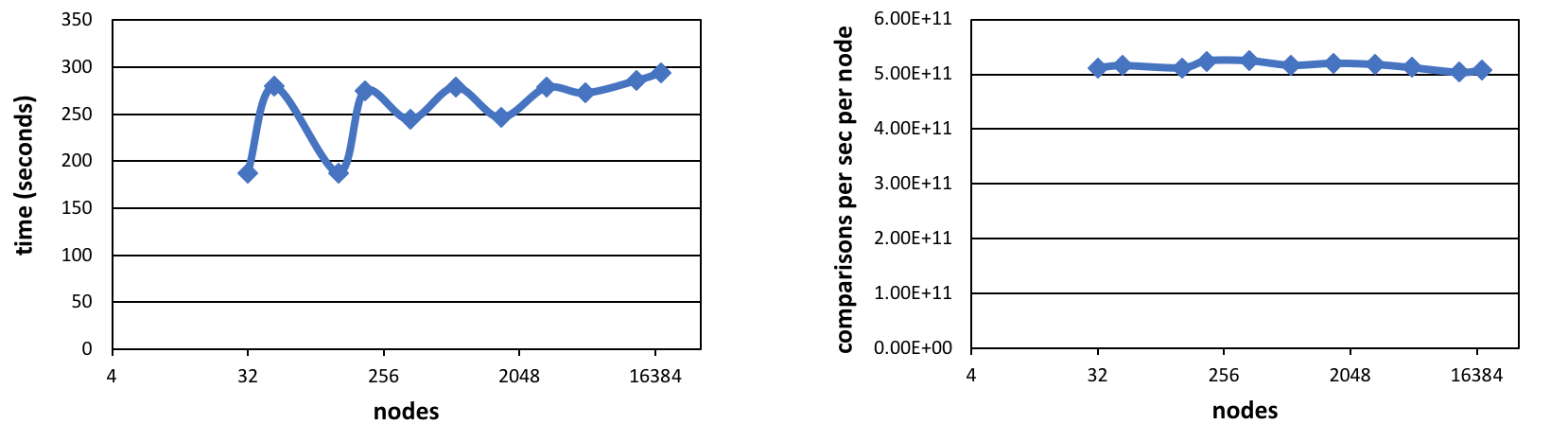}
\caption{CCC 2-way weak scaling. Left: time to solution. Right: number of unique elementwise comparisons per second per node.}
\label{fig:ccc-2way-results}
\end{figure}

Results are shown in Figure~\ref{fig:ccc-2way-results}.
The left graph shows good weak scaling timing performance up to the
full system.
The method benefits from the very high computational
intensity of the mGEMM2 kernel, dominating communication
costs which it asynchronously overlaps.
The right graph shows good weak scaling performance for the
comparison rate per node.
The maximum rate per node is 507e9 comparisons per second per node.
Here a comparison defined as the operation between
corresponding vector elements that produces four tally
values to be accumulated.
This rate is 87\% of the single node peak measured value of
582.8e9 from Table~\ref{tab:1gpu}, indicating almost perfect
efficiency at scale.

The method attains up to 9.11e15 comparisons
(9.11 petacomparisons) of two vector elements per second.
See Table~\ref{tab:ccc-comparisons-max}.


\subsection{3-way results}

The 3-way test runs are executed on up to
18,424 of Titan's 18,688 compute nodes, or 98.6\% of the system.
We use $n_f=20,000$ elements per vector and
$n_{vp}=n_v / n_{pv}=2,880$ vectors per node.
We set the load $\ell=6$, 
$n_{pr} = \lceil ( n_{pv} + 1 ) ( n_{pv} + 2 ) / \ell \rceil$,
and $n_{pf}=1$
and compute the final stage of $n_{st}=16$ stages.

\begin{figure}[ht]
\centering
\includegraphics*[height=1.5in]{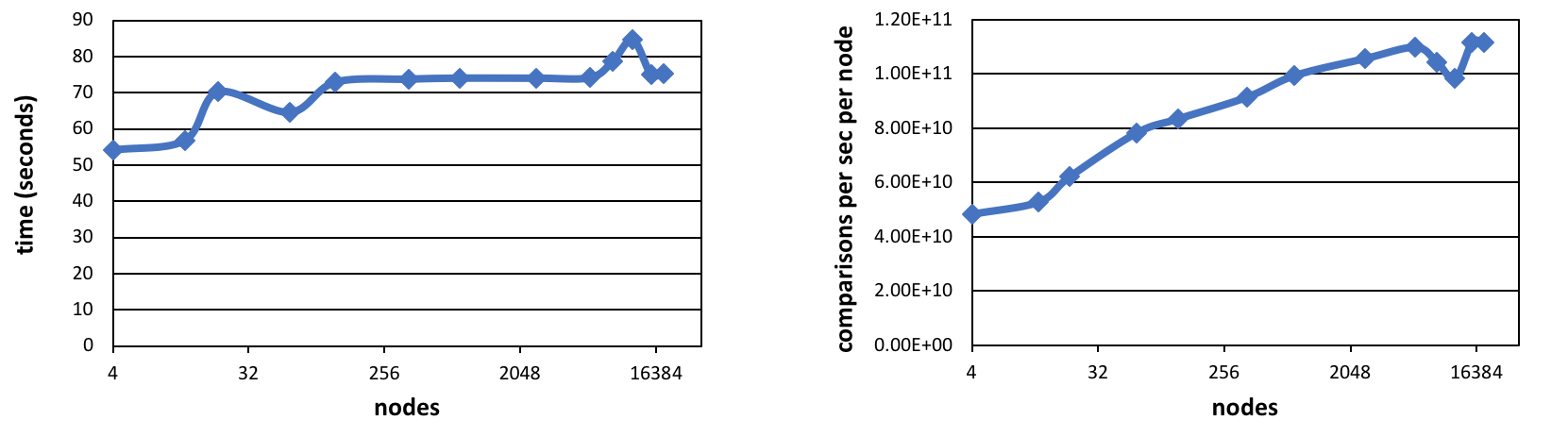}
\caption{CCC 3-way weak scaling. Left: time to solution. Right: number of unique elementwise comparisons per second per node.}
\label{fig:ccc-3way-results}
\end{figure}

Weak scaling results are shown in Figure~\ref{fig:ccc-3way-results}.
The left graph shows near ideal weak scaling timing
behavior for large problem sizes.
The right graph shows the average number of vector
element comparisons performed per second per node.
Here one comparison is defined as a single 3-way calculation
involving three vector elements that produces eight tallied results.
Some inefficiencies exist a lower node counts due to
known load balancing effects from the implementation;
the effects are minimal at large node counts.
The maximum rate per node is 112e9 comparisons per second;
this is 82\% of the peak single node measurable value of 136e9,
this value being 1/3 of the 4.077e11 mGEMM3 value from
Table~\ref{tab:1gpu} since three mGEMM3 operations are required
for each result.
Thus very high efficiency is achieved at scale.

Table~\ref{tab:ccc-comparisons-max} shows the maximum
operation and comparison rates attained at the highest node counts, up to
2.06 3-way petacomparisons per second.

\begin{table}
\begin{center}
\caption{Maximum performance, CCC}
\label{tab:ccc-comparisons-max}
\vspace{.05in}
\begin{tabular}{ccc}
method & comparisons \\
       & per second \\
\hline 
2-way CCC & $9.11\times 10^{15}$ \\
3-way CCC & $2.06\times 10^{15}$ \\
\end{tabular}
\end{center}
\end{table}


\subsection{Comparison with other work}

We now compare against results reported in the
literature as described in Section~\ref{section:introduction}.
We consider the most relevant comparable work we are aware of
as of this writing.
Table~\ref{tab:comparisons} shows comparisons per second
for alternative methods and implementations.
Unfortunately it is difficult to make rigorous comparisons
when architectures and algorithms are significantly different and
source code for the methods is not available or is not easily
ported and tuned to a single common architecture for comparison.
As a rough measure, a normalized performance ratio is calculated by normalizing
the (absolute) comparison rate
against the floating point rate of the respective hardware.
For newer hardware, the double precision peak rate is used;
for older GPUs with weak or nonexistent double precision support,
single precision is used.
This is used as a proxy for performance of bitwise integer operations
required for the core metrics calculations
on the respective processors,
as this information typically is not readily available.

We consider first the total number of comparisons per second.
Table~\ref{tab:comparisons} demonstrates that
the CoMet code used in the present work
gives several orders of magnitude higher absolute
performance, as measured in comparisons per second,
far beyond the demonstrated capability of any other code.

The normalized performance ratio of CoMet is similar to or better than
any result shown for other Kepler generation GPUs.
Normalized performance of CoMet is also within the general range of demonstrated
performance for cases using older Fermi GPUs, though it is not always the
fastest case by this rough measure.
It is unclear to what extent architectural differences between the GPUs
account for this.
Also, some other codes use a 3-bit representation of alleles rather
than the 2-bit representation used here;
this requires 50\% more storage but may result in higher performance than the
present implementation.
Additionally, the CPU results of \cite{pande} are far faster
than any other method, however this is because their algorithm
only compares single bits
rather than all combinations of 2 or 3 bits per element
as is done for the other cases.
A topic of future work will be to investigate further
performance optimizations for the present code.
In any case, it should be noted that
since CoMet has similar relative performance to codes like
\cite{gwisfi},
and since those codes already outperform conventional non-GPU
methods by whole orders of magnitude (see \cite{gwisfi} for details),
the same is necessarily true of CoMet.

\begin{table}
\begin{center}
\caption{Comparisons to related work}
\label{tab:comparisons}
\vspace{.05in}
\begin{tabular}{cccccccc}
code & problem & node config & nodes & GFlop rate & cmp/sec         & norm \\
     &         &             & used  &            & ($\times 10^9$) & perf \\
\hline 
\cite{pande}            & 2-way 1-bit& 1 Intel Core i7-920& 1 & 42.56 DP       & 222     &5.216 \\
GBOOST\cite{gboost}     & 2-way GWAS & 1 NVIDIA GTX 285   & 1 & 1062.72 SP     & 64.08   & .060 \\
GWISFI\cite{gwisfi}     & 2-way GWAS & 1 NVIDIA GTX 470   & 1 & 1088.6 SP      & 767     & .705 \\
\cite{goudey}           & 2-way GWAS & 1 NVIDIA GTX 470   & 1 & 1088.6 SP      & 649     & .596 \\
\cite{pande}            & 2-way 1-bit& 1 NVIDIA GTX 480   & 1 & 1345 SP        & 1185    & .881 \\
\cite{goudey}           & 2-way GWAS & IBM Blue Gene/Q    & 4096 & 839e3 DP    & 2520    & .003 \\
epiSNP\cite{weeks}      & 2-way GWAS & 2 Intel Xeon Phi SE10P & 126 & 271e3 DP & 1593    & .006 \\
\cite{gonzalez}         & 2-way GWAS & 2 NVIDIA K20m +    & 1 & 3360.56 DP     & 1053    & .313 \\
                        &            & 1 Intel Xeon Phi 5110P & &              &         &      \\
CoMet                   & 2-way CCC  & 1 NVIDIA K20X      & 17955 & 26.0e6 DP  & 9.110e6 & .350 \\
\hline 
GPU3SNP\cite{gonzalez2} & 3-way GWAS & 4 NVIDIA GTX Titan & 1 & 6000 DP        & 264.7   & .044 \\
CoMet                   & 3-way CCC  & 1 NVIDIA K20X      & 18424 & 26.7e6 DP  & 2.060e6 & .077 \\
\end{tabular}
\end{center}
\end{table}


\section{Future work}
\label{section:discussion}

The CoMet code here described will be used in a 2018
Department of Energy (DOE) INCITE project on the ORNL Titan
system, whose purpose is
to study genetic characteristics of \textit{Populus} species with
applications to production of cellulosic biomass fuels~\cite{incite2018}.
The transition of CoMet to production use for this and other
projects has necessitated ongoing development in the
following areas.

\begin{enumerate}

\item
Real-world genomic data is oftentimes sparse due to missing
data at some allele locations.
The approach described earlier
can be modified for this case in a straightforward fashion.
Since the input vector values $(0,1)$ and $(1,0)$ have an
identical effect on the calculated result, the value $(1,0)$
can be set aside as a marker to denote a missing entry in
the input vector.
The same approach described earlier using $(1,0)$ as a
marker for skipping a calculation for the 3-way case can
readily be adapted to skipping calculations for missing
entries in the sparse case.
Preliminary test show performance of the associated
modified GEMM kernel is only 9\% less
than the mGEMM3 rate shown in Table~\ref{tab:1gpu}.

\item
Our current model is designed for biallelic SNP data, in
which each marker has two states, either two nucleotide
states or insertion/deletion states, and we omit
multi-allelic SNPs. 2-way CCC in our current modeling returns four
values, representing the four possible combinations that a
biallelic variable can assume. In our future work, we plan
to extend our CCC measure to allow an arbitrary number of
marker states. When computing the correlation for a marker
with $s_1$ states and a marker with $s_2$ states, CCC will compute
the $s_1\times s_2$ possible relationships and return a matrix with
$s_1\times s_2$ values, at higher computational expense than
the current approach. The network model will have $s_1$
and $s_2$ nodes for the first and second markers, respectively. 

\item
To address instances of the 2-way case for which the results are too large
to fit into system memory, a ``phase'' technique can be
implemented, analogous to the staging technique
described earlier for the 3-way case.
Specifically, each processor's blocks of computed metrics results
are assigned phase numbers in a round-robin fashion
for a specified number of phases, and a single run of the
code will compute and store only a single phase of results
at a time.
The same approach can also reduce memory pressure for the 3-way case
in addition to the staging option.

\item
Past experience with other codes has shown that permuting the
process axis order for codes with multidimensional parallelism
may improve performance, by mapping processes with higher
communication bandwidth requirements to processors physically
closer in the system's interconnect topology.
In particular, for systems with multiple GPUs per node,
mapping the field processor axis (which has high parallel
reduction demands) to be in-node may be beneficial
to performance.

\item
The required genomics calculations are highly I/O intensive.
Experiments have shown that preprocessing the
input data into a single packed binary file and allowing each
compute node to read its required portion from that file
performs well in practice.

\item
Potentially vast quantities of output may be produced.
However in practice it
is often the case that very few of the elements are actually
needed---only those above a certain threshold size, which
may amount to less than one millionth of the calculated
values.
Tests have shown that in this regime it is efficient for
each compute node to write its results to a separate small
file.
To load balance the output, it is advantageous to
randomly permute the vectors of the input file as an
offline preprocessing step.
This removes correlations between
the parallel decomposition and the distribution of similar
vectors in the input data.
One small file per process should also be effective for burst
buffer hardware being delivered in newer HPC systems.

\item
It is likely that further code optimizations are possible,
for example, improving the performance of the mGEMM2/mGEMM3
operations and reducing other overheads.

\item
Porting to other architectures including
pre-exascale and exascale platforms is expected to be
tractable.
Support of Intel Xeon Phi nodes by modification of open source
dense linear algebra libraries is a possible option.

\end{enumerate}

Table~\ref{tab:summit}.
shows preliminary results with CoMet on the ORNL
Summit system~\cite{summit}.
In its final form Summit will be a 200 petaflop system with
4,608 compute nodes, each node composed of
two IBM Power9 processors and six NVIDIA Volta GPUs.
The experiments in Table~\ref{tab:summit}
apply the 2-way CCC method to a synthetic
dataset with 141,714 vectors of length 882 or 358,000 entries.
Comparisons with Titan are also shown, using 30 GPUs in both cases,
also with comparison to the corresponding CPU-only case.
The Summit vs.\ Titan per-GPU performance ratio for a large
problem that saturates both systems' GPUs is 6.44X,
slightly exceeding the roughly 5X peak flop rate ratio of the two
GPU models, most likely due to architectural improvements in the Volta GPU.
The smaller vector length 882 case does not saturate the GPU as
well; here the per-GPU performance of Summit exceeds Titan by a smaller factor.
The Summit GPU / CPU performance ratio is 119.2X, exceeding
the roughly 40X peak flop rate ratio between the six GPUs and the
two CPUs of each node, this being expected
since the CPU code version is not as heavily
optimized as the GPU version.
The largest Summit case shown here attains 3.328e12 comparisons per
second per GPU.
This rate extrapolated linearly to the eventual full Summit system yields
an estimate of 92.0 petacomparisons per second,
an unprecedented scale.

\begin{table}
\begin{center}
\caption{Summit vs.\ Titan performance, 30 GPUs, 30 Titan
nodes, 5 Summit nodes; runtimes in seconds}
\label{tab:summit}
\vspace{.05in}
\begin{tabular}{c|ccc||ccc}
\hline
system        & Titan & Summit & ratio & Titan  & Summit & ratio \\
vector length & 882   & 882   &       & 358,000 & 358,000 &  \\
\hline
GPU           & 5.63  & 1.86  & 3.02X &  232    & 36.0    & 6.44X\\
CPU           & ---   & 222.3 & ---   &  ---    & ---     & ---  \\
ratio         & ---   & 119.2X& ---   &  ---    & ---     & ---  \\
\end{tabular}
\end{center}
\end{table}



\section{Conclusions}
\label{section:conclusions}

We have defined a new set of algorithm implementations
for performing 2-way and 3-way Custom Correlation Coefficient
calculations for comparative genomics applications.
Performance of up to
nine quadrillion vector element comparisons per second is demonstrated.
To our knowledge this is the first simulation of its kind
ever performed at this scale,
demonstrating the capability to
perform simulations that were until recently considered
far beyond what is possible,
enabling new kinds of science in
comparative genomics.



\section*{Acknowledgments}

This research used resources of the Oak Ridge Leadership Computing
Facility at the Oak Ridge National Laboratory, which is supported by
the Office of Science of the U.S. Department of Energy under Contract
No. DE-AC05-00OR22725.

Funding provided by The BioEnergy Science Center (BESC) and The Center for Bioenergy Innovation (CBI).  U.S. Department of Energy Bioenergy Research Centers supported by the Office of Biological and Environmental Research in the DOE Office of Science.

Support for the Poplar GWAS dataset was provided by The BioEnergy Science (BESC) and The Center for Bioenergy Innovation (CBI). U.S. Department of Energy Bioenergy Research Centers supported by the Office of Biological and Environmental Research in the DOE Office of Science. The Poplar GWAS Project used resources of the Oak Ridge Leadership Computing Facility and the Compute and Data Environment for Science at Oak Ridge National Laboratory, which is supported by the Office of Science of the U.S. Department of Energy under Contract No. DE-AC05-00OR22725.

This research was also supported by the Plant-Microbe Interfaces Scientific Focus Area (http://pmi.ornl.gov) in the Genomic Science Program, the Office of Biological and Environmental Research (BER) in the U.S. Department of Energy Office of Science. Oak Ridge National Laboratory is managed by UT-Battelle, LLC, for the US DOE under contract DE-AC05-00OR22725.

This research was also supported by the Department of Energy Laboratory Directed Research and Development funding (7758), at the Oak Ridge National Laboratory. Oak Ridge National Laboratory is managed by UT-Battelle, LLC, for the US DOE under contract DE-AC05-00OR22725.

We would like to acknowledge Gerald Tuskan, Stephen DiFazio and the DOE Joint Genome Institute (JGI) for sequencing the \textit{Populus} genotypes and generating the processed SNP data.






\section*{References}

\bibliography{genomics_ccc_paper}


\end{document}